\documentclass[aps,twocolumn,prl]{revtex4}
\newcommand{\beq}{\begin{equation}} \newcommand{\eeq}{\end{equation}}
\newcommand{\bea}{\begin{eqnarray}} \newcommand{\eea}{\end{eqnarray}}
\newcommand{\bear}{\begin{eqnarray*}} \newcommand{\eear}{\end{eqnarray*}}
\newcommand{\lb}{\label} 
\newcommand{\rf}[1]{(\ref{#1})}
\usepackage[pdftex]{graphicx}
\usepackage{epstopdf}
\usepackage{amsmath}
\usepackage{amsfonts}
\usepackage{amssymb}
\usepackage{makeidx}

\begin{document}

\title {Asymmetric exclusion model with impurities.}

\author{Matheus J. Lazo$^{1}$\footnote{matheuslazo@furg.br} and Anderson A. Ferreira$^2$}

\affiliation{$^1$ Instituto de Matem\'atica, Estat\'\i stica e F\'\i sica - FURG, Rio Grande, RS, Brazil.\\$^2$ Instituto de F\'{\i}sica Te\'orica, Unesp, S\~ao Paulo, SP, Brazil.}

\begin{abstract}

An integrable asymmetric exclusion process with impurities is formulated. The model displays the full spectrum of the stochastic asymmetric XXZ chain plus new levels. We derive the Bethe equations and calculate the spectral gap for the totally asymmetric diffusion at half filling. While the standard asymmetric exclusion process without impurities belongs to the KPZ universality class with a exponent $\frac{3}{2}$, our model has a scaling exponent $\frac{5}{2}$.


\end{abstract}

\maketitle

%%%%%%%%%%%%%%%%%%%%%%%%%%%%%%%%%%%%%%%%%%%%%%%%%%%%%%%%%%%%%%%%%%%%%%%%%%%%%%%%%%%%%%%%%%%%

One-dimensional $3$-state quantum Hamiltonians and master equations of reaction-diffusion processes have played an important role in describing strongly correlated electrons and non-equilibrium statistical mechanics in the last decades, mainly due to their intrinsic and nontrivial many-body behavior. Remarkably, in one dimension several models in this category are exactly solvable, as the spin-$1$ Sutherland \cite{Sutherland} and $t$-$J$ \cite{schlo} models, and the asymmetric diffusion of two types of particles \cite{stochastic}. In its formulation in terms of particles with two global conservation laws, these models describe the dynamics of two types of particles on the lattice, where the total number of particles of each type is conserved separately. In order to ensure integrability, all known models in this class satisfies some particle-particle exchange symmetries \cite{popkov,lazo2}. Recently, we introduced a new class of $3$-state model in the context of high energy physics that is integrable despite it do not have particle-particle exchange symmetry \cite{lazo2}. The quantum version is closely related to the strong regime of the $t$-$U$ Hubbard model and the XXC model \cite{chico}. In this work we formulate a stochastic model related to \cite{lazo2} that describes an asymmetric exclusion process (ASEP) with impurities. Although our model can be solved by the coordinate Bethe ansatz, we are going to formulate a new matrix product ansatz (MPA) \cite{alclazo2,popkov} due its simplicity and unifying implementation for arbitrary systems. This new MPA introduced in \cite{alclazo2,popkov} can be seen as a matrix product formulation of the coordinate Bethe ansatz and it is suited to describe all eigenstates of integrable models. We solve this model with periodic boundary condition through the MPA and we analyze the spectral gap for some special cases. Our model displays the full spectrum of the ASEP without impurities \cite{GwaSpohn} plus new levels. The first excited state belongs to these new levels and has unusual scaling exponents. Although the ASEP without impurities belongs to the KPZ universality class \cite{KPZ} (dynamic exponent $\frac{3}{2}$), our new model displays a scaling exponent $\frac{5}{2}$.

The model we propose describes the dynamics of two types of particles (type $1$ and $2$) on the lattice, where the total number of particles of each type is conserved. In this model if the neighbors sites are empty, particles of type $1$ can jump to the right or to the left with rate $\Gamma_{0\;1}^{1\;0}$ and $\Gamma_{1\;0}^{0\;1}$, respectively. Particles of type $2$ ``impurities'' do not jump to the neighbors sites if they are empty, but can change positions with neighbors particles of type $1$ with rates $\Gamma_{2\;1}^{1\;2}$ and $\Gamma_{1\;2}^{2\;1}$ if the particle $1$ is on the left or on the right, respectively. There are not particle-particle exchange symmetry because particles $1$ and $2$ have different kinds of dynamics. The stochastic model we consider on a ring of perimeter $L$ is related to the Hamiltonian $H=H^{(0)}+H^{(2)}$ where ($\alpha=0,2$)
\bea
\lb{e1}
H^{(\alpha)}&=&\sum_{j=1}^{L}\left[\Gamma_{\alpha\;1}^{1\;\alpha} (E_j^{\alpha,1}E_{j+1}^{1,\alpha}-E_j^{1,1}E_{j+1}^{\alpha,\alpha})\right. \\
&&\;\;\;\;\;\;\;\;\;\;\;\;\;\;\;\left.+\Gamma_{1\;\alpha}^{\alpha\;1} (E_j^{1,\alpha}E_{j+1}^{\alpha,1}-E_j^{\alpha,\alpha}E_{j+1}^{1,1})\right.,\nonumber
\eea
and $E^{\alpha,\beta}$ ($\alpha,\beta=0,1,2$) are the usual $3\times 3$ Weyl matrices with $i,j$ elements $\left(E^{l,m} \right)_{i,j}=\delta_{l,i}\delta_{m,j}$, and $\Gamma_{\alpha\;1}^{1\;\alpha}+\Gamma_{1\;\alpha}^{\alpha\;1}=1$. Finally, the model \rf{e1} is related to the tracer diffusion model \cite{popkov} by an interchange of labels $0$ and $1$, however the exactly solutions of these models describe different sectors of particles occupation.

The conservation of particles supplemented by the periodic boundary condition of \rf{e1} imply that the total number of particles $n_1,n_2$ on class $1$ and $2$ as well the momentum $P=\frac{2\pi l}{L}$ ($l=0,1,\ldots,L-1$) are good quantum numbers. The eigenstates $|\Psi\rangle$ of the eigenvalue equation $H|\Psi\rangle=\varepsilon^{n_1,n_2}|\Psi\rangle$ belonging to the eigensector labeled by ($n_1,n_2,P$) are given by
\beq
|\Psi\rangle=\sum_{\{\alpha\},\{x \}} f(x_1,\alpha_1;\ldots;x_n,\alpha_n)|x_1,\alpha_1;\ldots;x_n,\alpha_n\rangle,
\lb{e3}
\eeq
where the kets $|x_1,\alpha_1;\ldots;s_n,\alpha_n\rangle$ denote the configurations with particles of type $\alpha_i$ ($\alpha_i=1,2$) located at the position $x_i$ ($x_i=1,\ldots,L$) and $n=n_1+n_2$ is the total number of particles. The summation $\{\alpha \}=\{\alpha_1,\ldots,\alpha_n\}$ extends over all the permutations of $n$ numbers $\{1,2 \}$ in which $n_1$ terms have value $1$ and $n_2$ terms the value $2$, while the summation $\{x \}=\{x_1,\ldots,x_n \}$ extends, for each permutation $\{\alpha \}$, into the set of the non-decreasing integers satisfying $x_{i+1}\ge x_i+1$.

The MPA asserts that the amplitudes of an arbitrary eigenfunction \rf{e3} can be mapped to a matrix product
\bea
\lb{e4}
&&\!\!\!\!\!f(x_1,\alpha_1;\ldots;x_n,\alpha_n)\Longleftrightarrow\\
&&\!\!\!\!\!E^{x_1-1}A^{(\alpha_1)}E^{x_2-x_1-1}A^{(\alpha_2)}\cdots E^{x_n-x_{n-1}-1}A^{(\alpha_n)}E^{L-x_n},\nonumber
\eea
where for this map we can choose any operation on the matrix products that gives a non-zero $c$-number \cite{alclazo2}.
The matrices $A^{(\alpha)}$ are associated to the particles of type $\alpha$ $(\alpha=1,2)$ and the matrix $E$ is associated to the vacant sites \cite{alclazo2}. Since the eigenfunction \rf{e3} has a well defined momentum $P=\frac{2\pi l}{L}$ due to the periodic boundary condition, the matrix product \rf{e4} should satisfy the constraints $f(x_1,\alpha_1;\ldots;x_n,\alpha_n)=e^{-iP}f(x_1+1,\alpha_1;\ldots;x_n+1,\alpha_n)$ for $x_n<L$, and $f(x_1,\alpha_1;\ldots;x_n,\alpha_n)=e^{-iP}f(1,\alpha_n;\ldots;x_{n-1}+1,\alpha_{n-1})$ for $x_n=L$. Let us consider initially the simple cases where $n=1$ and $n=2$.

%%%%%%%%%%%%%%%%%%%%%%%%%%%%%%%%%%%%%%%%%%%%%%%%%%%%%%%%%%%%%%%%%%%%%%%%%%%%%%%%%%%%%%%%%%%%%%%%%%%

{\it {\bf  n = 1.}}
Since the Hamiltonian \rf{e1} is diagonal for $n_1=0$, let us consider only the case $n_1=1$. By inserting the ansatz \rf{e4} in \rf{e3}, the eigenvalue equation gives us
\bea
\lb{e5}
&&\varepsilon^{1,0} E^{x-1}A^{(1)}E^{L-x}=\Gamma_{0\;1}^{1\;0} E^{x-2}A^{(1)}E^{L-x+1} \\
&&+\Gamma_{1\;0}^{0\;1} E^{x}A^{(1)}E^{L-x-1} - E^{x-1}A^{(1)}E^{L-x}, \nonumber
\eea
A convenient solution of \rf{e5} is obtained by introducing the spectral parameter dependent matrix $A^{(1)}=EA_k^{(1)}$ with $EA_k^{(1)}=e^{ik}A_k^{(1)}E$ and $k \in \mathbb{C}$. Inserting these relations into \rf{e5} we obtain
\beq
\varepsilon(k)\equiv \varepsilon^{1,0}=\Gamma_{0\;1}^{1\;0}e^{-ik}+\Gamma_{1\;0}^{0\;1}e^{ik}-1.
\lb{e9}
\eeq
The up to now free spectral parameter $k$ is fixed by imposing the boundary condition. This will be done only for the general $n$.

%%%%%%%%%%%%%%%%%%%%%%%%%%%%%%%%%%%%%%%%%%%%%%%%%%%%%%%%%%%%%%%%%%%%%%%%%%%%%%%%%%%%%%%%%%%%%%%%%%%

{\it {\bf  n = 2.}}
For two particles of types $\alpha_1$ and $\alpha_2$ ($\alpha_1,\alpha_2=1,2$) we have two kinds of relations coming from the eigenvalue equation. The configurations where the particles are at positions ($x$, $y$) with $y-x=d>1$ give us the generalization of \rf{e5}, and the configurations where the particles are at the colliding positions ($y=x+1$) give us
\bea
&&\varepsilon^{\alpha_1,\alpha_2} EA^{(\alpha_1)}A^{(\alpha_2)}E= \Gamma_{0\;\alpha_1}^{\alpha_1\;0}A^{(\alpha_1)}EA^{(\alpha_2)}E \nonumber \\
&&+\Gamma_{\alpha_2\;0}^{0\;\alpha_2}EA^{(\alpha_1)}EA^{(\alpha_2)} + \Gamma_{\alpha_1\;\alpha_2}^{\alpha_2\;\alpha_1}EA^{(\alpha_2)}A^{(\alpha_1)}E \nonumber \\
&&-\left(\Gamma_{\alpha_1\;0}^{0\;\alpha_1}+\Gamma_{0\;\alpha_2}^{\alpha_2\;0}+\Gamma_{\alpha_2\;\alpha_1}^{\alpha_1\;\alpha_2}\right) EA^{(\alpha_1)}A^{(\alpha_2)}E,
\lb{e11}
\eea
where we introduced $\Gamma_{1\;1}^{1\;1}=\Gamma_{2\;2}^{2\;2}= \Gamma_{0\;2}^{2\;0}=\Gamma_{2\;0}^{0\;2}=0$ and $\varepsilon^{0,2}=0$. A solution of these relations is obtained by identifying $A^{(\alpha)}$ as composed by $n_{\alpha}$ spectral parameter dependent matrices $A_{k_1^{(\alpha)}}^{(\alpha)}$ and $A_{k_2^{(\alpha)}}^{(\alpha)}$ \footnote{It is important to mention that as we have two distinct sets of spectral parameters, different from the models in \cite{chico}, our model will not have a $R$ matrix on the difference of spectral parameters.}, i. e.,
\beq
A^{(\alpha)}=\sum_{j=1}^{n_\alpha}EA_{k_j^{(\alpha)}}^{(\alpha)}, \;\; {\mbox{with}} \;\; EA_{k_j^{(\alpha)}}^{(\alpha)}=e^{ik_j^{(\alpha)}}A_{k_j^{(\alpha)}}^{(\alpha)}E, 
\lb{e12}
\eeq
with $\alpha=1,2$ and $j=1,..., n_\alpha$. By inserting \rf{e12} in the first kind of relations, where $y-x=d>1$, we obtain the energy in terms of the spectral parameters
\beq
\varepsilon^{n_1,n_2}=\sum_{j=1}^{n_1}\varepsilon(k_j^{(1)}).
\lb{e32}
\eeq
where $\varepsilon(k)$ is given by \rf{e9}. Let us consider now \rf{e11} in the case where the particles are of same type. Using \rf{e12} and \rf{e32} in \rf{e11} for $\alpha_1=\alpha_2=1$ we obtain the relations 
\beq
A_{k_j^{(\alpha_1)}}^{(\alpha_1)}A_{k_l^{(\alpha_2)}}^{(\alpha_2)}=S_{\alpha_1\;\alpha_2}^{\alpha_1\;\alpha_2}(k_j^{(\alpha_1)},k_l^{(\alpha_2)})A_{k_l^{(\alpha_2)}}^{(\alpha_2)}A_{k_j^{(\alpha_1)}}^{(\alpha_1)},
\lb{e33}
\eeq
where $(A_{k_j^{(\alpha_j)}}^{(\alpha_j)})^2=0$ and
\beq
S_{1\;1}^{1\;1}(k_j^{(1)},k_l^{(1)})=-\frac{\Gamma_{0\;1}^{1\;0}+\Gamma_{1\;0}^{0\;1}e^{i(k_j^{(1)}+k_l^{(1)})}- e^{ik_j^{(1)}}}{\Gamma_{0\;1}^{1\;0}+\Gamma_{1\;0}^{0\;1}e^{i(k_j^{(1)}+k_l^{(1)})}-e^{ik_l^{(1)}}}.
\lb{e15}
\eeq
For two impurities at "colliding" positions, the eigenvalue equation does not fix a commutation relation between the matrices $A_{k_j^{(2)}}^{(2)}$. On the other hand, the sum
\beq
\sum_{j,l=1}^2 A_{k_j^{(2)}}^{(2)}E^dA_{k_l^{(2)}}^{(2)}\neq 0,
\lb{e16}
\eeq
where the number of vacant sites between the impurities $d=y-x$ is a conserved charge of the Hamiltonian \rf{e1}, needs to be different from zero or the MPA will produces an eigenfunction with null norm. Moreover, the algebraic expression in \rf{e12} assures that any matrix product defining our ansatz \rf{e4} can be expressed in terms of two single matrix products $A_{k_1^{(2)}}^{(2)}A_{k_2^{(2)}}^{(2)}E^L$ and $A_{k_2^{(2)}}^{(2)}A_{k_1^{(2)}}^{(2)}E^L$. Using \rf{e12} we have, from the periodic boundary condition,
\beq
A_{k_j^{(2)}}^{(2)}A_{k_l^{(2)}}^{(2)}E^L=e^{-ik_j^{(2)}L}e^{-ik_l^{(2)}L}A_{k_j^{(2)}}^{(2)}A_{k_l^{(2)}}^{(2)}E^L,
\lb{e19}
\eeq
To satisfy this equation we should have $k_2^{(2)}=-k_1^{(2)} + 2\pi j/L$ ($j=0,1,...,L-1$). Consequently, from \rf{e19} the most general commutation relation among the matrices $A_{k_1^{(2)}}^{(2)}$ and $A_{k_2^{(2)}}^{(2)}$ can be reduced to $A_{k_1^{(2)}}^{(2)}A_{k_2^{(2)}}^{(2)}=A_{k_2^{(2)}}^{(2)}A_{k_1^{(2)}}^{(2)}$ (or $S_{2\;2}^{2\;2}(k_j^{(2)},k_l^{(2)})=1$ in \rf{e33}) by an appropriate change of variable in the spectral parameter $k_1^{(2)}$. Inserting these results in the sum of \rf{e16} and by imposing that the sum is not zero, we obtain ($j,l,v=1,...,n_2$)
\beq
k_l^{(2)}\neq k_j^{(2)} + \frac{\pi (2m+1)}{d_v} \;\; (m=0,1,...),
\lb{e23}
\eeq
where $\{d_v\}$ is the set of all numbers of vacant sites between the impurities.

Let us consider now the case where the particles are of distinct species. By using \rf{e9}, \rf{e32} and \rf{e12}, the equation \rf{e11} gives us two independent relations for $\alpha_1 \ne \alpha_2=1,2$:
\bea
\lb{e27}
&&\!\!\!\!\! \left[\Gamma_{0\;\alpha_2}^{\alpha_2\;0}+\Gamma_{\alpha_1\;0}^{0\;\alpha_1}e^{i(k^{(\alpha_1)}+k^{(\alpha_2)})}-t_{\alpha_1 \alpha_2}e^{ik^{(\alpha_2)}} \right]A_{k^{(\alpha_1)}}^{(\alpha_1)}A_{k^{(\alpha_2)}}^{(\alpha_2)} \nonumber \\
&&\;\;\;\;\;\;\;\;\;\;\;\;\;\;\;\;\;\;\;\;\;\;\;\;=\Gamma_{\alpha_1\;\alpha_2}^{\alpha_2\;\alpha_1}e^{ik^{(\alpha_2)}}A_{k^{(\alpha_2)}}^{(\alpha_2)}A_{k^{(\alpha_1)}}^{(\alpha_1)},
\eea
where $t_{\alpha_1 \alpha_2}=1-\Gamma_{\alpha_1\;0}^{0\;\alpha_1}-\Gamma_{0\;\alpha_2}^{\alpha_2\;0}-\Gamma_{\alpha_2\;\alpha_1}^{\alpha_1\;\alpha_2}$. This two relations need to be identically satisfied since at this level we want to keep $k^{(\alpha)}$ as free complex parameters. Then \rf{e27} imply special choices of the coupling constants $\Gamma_{k\;l}^{m\;n}$:
\beq
\lb{e28}
\Gamma_{2\;1}^{1\;2}=\Gamma_{0\;1}^{1\;0},\;\;\;\; \Gamma_{1\;2}^{2\;1}=\Gamma_{1\;0}^{0\;1}.
\eeq
The integrability conditions \rf{e28} reduces the $2$ parameters defining the Hamiltonian \rf{e1} into only $1$ free parameter $\frac{\Gamma_{0\;1}^{1\;0}}{\Gamma_{1\;0}^{0\;1}}$. Moreover in order to the model be integrable, the particle moves over impurities as if the impurities are vacant sites. This condition will play a fundamental role in the spectral properties of the Hamiltonian \rf{e1}, giving us the full spectrum of the standard ASEP plus new levels. From \rf{e27} and \rf{e28} we obtain the structural constants:
\beq
\lb{e29}
S^{2 \; 1}_{2 \; 1}(k^{(2)},k^{(1)})=\frac{1}{S^{1 \; 2}_{1 \; 2}(k^{(1)},k^{(2)})}=e^{ik^{(2)}}.
\eeq

%%%%%%%%%%%%%%%%%%%%%%%%%%%%%%%%%%%%%%%%%%%%%%%%%%%%%%%%%%%%%%%%%%%%%%%%%%%%%%%%%%%%%%%%%%%%%%%%%%%

{\it {\bf  General n.}}
We now consider the case of arbitrary numbers $n_1$, $n_2$ of particles of type $1$ and $2$. The eigenvalue equation gives us generalizations of \rf{e5} and \rf{e11}. To solve these equations we identify the matrices $A^{(\alpha)}$ as composed by $n_{\alpha}$ spectral dependent matrices \rf{e12}. The configurations where $x_{i+1}>x_i+1$ give us the energy \rf{e32}. The amplitudes in \rf{e3} where a pair of particles of types $\alpha_1$ and $\alpha_2$ are located at the closest positions give us the algebraic relations \rf{e33} where the algebraic structure constants are the diagonal $S$-matrix defined by \rf{e15}, \rf{e29}, $S_{2\;2}^{2\;2}(k_j^{(2)},k_l^{(2)})=1$, with coupling constants \rf{e28}.

In order to complete our solutions we should fix the spectral parameters ($k_1^{(1)},\ldots, k_{n_1}^{(1)}$ and $k_1^{(2)},\ldots, k_{n_2}^{(2)}$). Using the algebraic relations \rf{e33} for an arbitrary amplitude $A_{k_1^{(1)}}^{(1)}\cdots  A_{k_{n_1}^{(1)}}^{(1)}A_{k_1^{(2)}}^{(2)}\cdots A_{k_{n_2}^{(2)}}^{(2)} E^L$ we obtain, from the periodic boundary condition,
\bea
&&e^{ik_j^{(1)}L}=-e^{-i\sum_{l=1}^{n_2}k_l^{(2)}}\prod_{l=1}^{n_1}S_{1\;1}^{1\;1}(k_j^{(1)},k_l^{(1)})\nonumber \\
&&e^{ik_q^{(2)}(L-n_1)}=1,
\lb{e36}
\eea
where we have $(j=1,...,n_1)$ and $(q=1,...,n_2)$. 

Finally, the momentum of the eigenstate is given by inserting the ansatz \rf{e12} into the relation  $f(x_1,\alpha_1;\ldots;x_n,\alpha_n)=e^{-iP}f(x_1+1,\alpha_1;\ldots;x_n+1,\alpha_n)$:
\beq
\lb{e35b}
P=\sum_{j=1}^{n_1} k_j^{(1)} + \sum_{j=1}^{n_2} k_j^{(2)}=\frac{2\pi l}{L} \;\;\; (l=0,1,\ldots,L-1).
\eeq

%%%%%%%%%%%%%%%%%%%%%%%%%%%%%%%%%%%%%%%%%%%%%%%%%%%%%%%%%%%%%%%%%%%%%%%%%%%%%%%%%%%%%%%%%%%%%%%%%%%
{\bf Spectral Gap.} We can write the Bethe equation \rf{e36} in a more convenient way. From the second expression on \rf{e36} we have
\beq
e^{-i\sum_{j=1}^{n_2}k_j^{(2)}}=e^{-i\frac{2\pi m}{L-n1}}=\phi_m
\lb{e38}
\eeq
with $m=0,1,...,L-n_1-1$. By inserting \rf{e38} and using \rf{e15} in the first equation in \rf{e36} we obtain
\beq
e^{ik_jL}=(-)^{n_1-1}\phi_m\prod_{l=1}^{n_1}\frac{\Gamma_{0\;1}^{1\;0}+\Gamma_{1\;0}^{0\;1}e^{i(k_j+k_l)}- e^{ik_j}}{\Gamma_{0\;1}^{1\;0}+\Gamma_{1\;0}^{0\;1}e^{i(k_j+k_l)}- e^{ik_l}},
\lb{e39}
\eeq
where $j =1,...,n_1$ and $k_j\equiv k_j^{(1)}$. The Bethe equation \rf{e39} differs from the one related to the stochastic version of the asymmetric XXZ chain \cite{GwaSpohn} by the phase factor $\phi_m$ defined in \rf{e38}. However, different from the phase factor in the asymmetric XXZ with long range interactions \cite{AlcBar}, the phase $\phi_m$ defined in \rf{e38} plays a fundamental role in the spectral properties of the model by changing the scaling behavior of the system. While in \cite{AlcBar} the phase is a constant, in our model it is a function on the momentum of the state through the relation \rf{e35b}. For $m=0$ we obtain the full spectrum of the standard ASEP \cite{GwaSpohn}. Additional energy levels are given by different choices for $m$, and the first excited state belongs to these new levels. 

Finally, it is important to notice that the Bethe equation \rf{e39} do not depends on the number of impurities $n_2\neq 0$. Consequently, the roots of \rf{e39} and the eigenvalues \rf{e32} are independent of $n_2\neq 0$ (but the wave function depends on $n_2$). This huge spectrum degeneracy follows from the conservation of the numbers of vacant sites between the impurities by the Hamiltonian \rf{e1}. Let us explain with the following example. Suppose we start with the given configuration $0120220$ with one particle ($1$), $3$ impurities ($2$) and $3$ vacant sites ($0$). We can make a surjective map between all possible configuration of these particles to all possible configurations of a new chain with just impurities and vacant sites. For example $0120220 \Longrightarrow 020220$. On this new chain, we are looking only for the effective movement of impurities on the chain. For simplicity, let us consider the totally asymmetric model (TASEP) where $\Gamma_{0\;1}^{1\;0}=1$ and $\Gamma_{1\;0}^{0\;1}=0$. When the particle jumps over the impurities, nothing changes in the effective chain since we also have $0210220 \Longrightarrow 020220$, then $0201220 \Longrightarrow 020220$, then $0202120 \Longrightarrow 020220$, then $0202210 \Longrightarrow 020220$, then $0202201 \Longrightarrow 020220$, and finally a change in the mapped configuration $1202200 \Longrightarrow 202200$. On other words, the impurities move on the mapped chain as they are just one "object" due to the conservation of vacant sites between impurities. Moreover, this "object" only moves when the particle complete a turn over the chain. As a consequence, the time for the particle to complete one revolution is the time scale for the movements of the this "object". For an arbitrary number of particles in a chain of length $L$, the time for the particles to complete one revolution is of order $L^{\frac{3}{2}}$. As the "object" formed by the impurities need to moves of order $L$ times to span all possible configurations, we will need a time of order $L^{\frac{3}{2}}\times L = L^{\frac{5}{2}}$ to reach the stationary state. 

The Bethe equation \rf{e39} for the TASEP can be written as \cite{GwaSpohn} 
\beq
(1-z_j)^{n_1}\left(1+z_j\right)^{L-n_1}=-2^{L}\phi_m\prod_{l=1}^{n_1}\frac{z_l-1}{z_l+1}\equiv Y,
\lb{e45}
\eeq
where $z_j=2e^{-ik_j}-1$, and we introduced an auxiliary complex variable $Y$. It can be noted that the right-hand side of these equations is independent of the index $j$, consequently the Bethe equation \rf{e45} can be reduced to a simple polynomial equation of degree $L$ as in \cite{GwaSpohn}. In the half-filling sector $n_1=L/2$ this polynomial is further simplified to $Y=\left(1-z^2\right)^{n_1}$. To solve this Bethe equation we follow the method developed in \cite{GwaSpohn} and we pick a $Y$ and we take its $n_1th$ root. These roots give us $2n_1$ solutions $z_j$. For each particular choice of $n_1$ out of $2n_1$ solutions we have to determine $Y$ through \rf{e45}. On the other hand, we also solve \rf{e45} numerically by direct computation of the roots of \rf{e45} up to $L=1024$. The stationary state is provided by choosing $m=0$ and $P=0$ and selecting the $n_1$ fugacities $z_j$ with the largest real parts. The first excited state is also obtained by selecting the $n_1$ fugacities with the largest real parts and by choosing $m=1$ and $P=\frac{2\pi}{L}$. The energy gap has a leading behavior of KPZ $L^{-\frac{3}{2}}$ and a sub-leading term $L^{-\frac{5}{2}}$ related to the super-diffusion of particles $1$ and sub-diffusion of particles $2$, respectively. For $m=0$ (and also for $n_2=0$) the sub-leading term vanishes and we recover the spectrum of the ASEP without impurities \cite{GwaSpohn}. For $m=1$ the spectral gap scales with $L^{-\frac{5}{2}}$ instead of $L^{-\frac{3}{2}}$ due to the sub-leading term. It is important to notice that the dynamics of particles $1$ is not affected by the impurities. This explain why the model displays the full spectrum of the ASEP. On the other hand, the dynamics of the impurities is totally dependent on the particles, since the impurities only move when particles change position with them. Consequently, the time to vanish the fluctuations on the densities of particle acts as a time scale for the diffusion of the impurities, resulting in a relaxation time longer than the one for the standard ASEP and reflected in the $L^{-\frac{3}{2}}\times L^{-1} = L^{-\frac{5}{2}}$ gap.

{\bf Eigenstates.} The MPA \rf{e4} enable us to write all eigenstates of \rf{e1} in a matrix product form. For a given solution $k_j$ and $k_j^{(2)}$, the matrices $E$ and $A_{k_j^{(\alpha)}}^{(\alpha)}$ have the following finite-dimensional representation:
\bea
\lb{e46}
E&=&\bigotimes_{l=1}^{n_1} \left( \begin{array}{cc}
				  1 & 0 \\
                                  0 & e^{ik_l}
			          \end{array}
			  	 \right)
				  \bigotimes_{l=1}^{n_2} \left( \begin{array}{cc}
				  1 & 0 \\
                                  0 & e^{ik^{(2)}_l}
			          \end{array}
			          \right), \\
A^{(2)}_{k^{(2)}_j}&=&\bigotimes_{l=1}^{n_1} \left( \begin{array}{cc}
				  1 & 0 \\
                                  0 & e^{ik^{(2)}_l}
			          \end{array}
			  \right)
		    \bigotimes_{l=1}^{j-1} I_2\bigotimes
				 \left( \begin{array}{cc}
				  0 & 0 \\
                                  1 & 0
			          \end{array}
			  \right)
			\bigotimes_{l=j+1}^{n_2} I_2,\nonumber \\
A^{(1)}_{k_j}&=&\bigotimes_{l=1}^{j-1} \left( \begin{array}{cc}
				  S^{1 \; 1}_{1 \; 1}(k_j,k_l) & 0 \\
                                  0 & 1
			          \end{array}
			  \right)\bigotimes\left( \begin{array}{cc}
				  0 & 0 \\
                                  1 & 0
			          \end{array}
			  \right)
		    \bigotimes_{j=j+1}^{n_1+n_2} I_2,
\nonumber
\eea
where $n=n_1+n_2$, $I_2$ is the $2\times 2$ identity matrix, and the dimension of the representation is $2^n$. The matrix product form of the eigenstates are given by inserting the matrices \rf{e12} defining the MPA \rf{e4} with the spectral parameters \rf{e39} into equation \rf{e3}. As showed in \cite{mallick}, our MPA generalizes the steady-state Matrix Product introduced by Derrida et al \cite{Derrida}. The stationary state is obtained by choosing $k_j=0$ and $k_j^{(2)}=0$ in \rf{e46}. However, a relation between our matrix product form for the steady-state and the standard Matrix Product form \cite{Derrida} is not trivial \cite{mallick}. 

Finally, like in the ASEP, the stationary state is obtained by a combination of all possible configuration of particles and impurities with the same probability (satisfying the order $\{d_1,...,d_{n_2}\}$ of vacant sites between impurities, up to cyclic permutations). This can be seen from the fact that the particles move on the chain as if the impurities are vacant sites and the impurities moves on the surjective mapped chain as a single "object".

The spectral gap for other sectors and other boundary conditions and the physical consequences of these, are currently under investigation.

We are grateful to G.M. Sch\"utz and F. C. Alcaraz by the very helpful discussions.

%%%%%%%%%%%%%%%%%%%%%%%%%%%%%%%%%%%%%%%%%%%%%%%%%%%%%%%%%%%%%%%%%%%%%%%%%%%%%%%%%%%%%%%%%%

\end{document}